\documentclass{aa}
\usepackage{graphicx}
\usepackage[varg]{txfonts}
\usepackage{cases}
\usepackage{natbib}

\newcommand{\mgii}{Mg \scriptsize{II} \normalsize}
\newcommand{\cii}{C \scriptsize{II} \normalsize}
\newcommand{\caii}{Ca \scriptsize{II} \normalsize}
\newcommand{\hei}{He \scriptsize{I} \normalsize}

\begin{document}

\title{Doppler shift oscillations of a sunspot detected by CYRA and IRIS}

\author{D.~Li \inst{1,2,3}, X.~Yang\inst{4,6}, X.~Y.~Bai\inst{2}, J.~T.~Su\inst{2,5}, Z.~J.~Ning\inst{1}, W.~Cao\inst{4,6} and Y.~Y.~Deng\inst{2,5}}

\institute{Key Laboratory for Dark Matter and Space Science, Purple Mountain Observatory, CAS, Nanjing 210033, PR China \email{lidong@pmo.ac.cn} \\
           \and CAS Key Laboratory of Solar Activity, National Astronomical Observatories, Beijing 100012, PR China \email{xybai@bao.ac.cn} \\
           \and State Key Laboratory of Space Weather, Chinese Academy of Sciences, Beijing 100190, PR China \\
           \and Big Bear Solar Observatory, New Jersey Institute of Technology, Big Bear City, CA 92314-9672, USA \\
           \and School of Astronomy and Space Sciences, University of Chinese Academy of Sciences, Beijing 100049, PR China \\
           \and Center for Solar-Terrestrial Research, New Jersey Institute of Technology, 323 Martin Luther King Boulevard, Newark, NJ 07102, USA \\
           }
\date{Received; accepted}

\titlerunning{Doppler shift oscillations of a sunspot detected by CYRA and IRIS}
\authorrunning{D.~Li et al.}

\abstract {The carbon monoxide (CO) molecular line at around
46655~{\AA} in solar infrared spectra is often used to investigate
the dynamic behavior of the cold heart of the  solar atmosphere,
i.e., sunspot oscillation, especially at the sunspot umbra.} {We
investigated  sunspot oscillation at Doppler velocities of the
CO~7-6~R67 and 3-2~R14 lines that were measured by the Cryogenic
Infrared Spectrograph (CYRA), as well as the line profile of \mgii~k
line that was detected by the Interface Region Imaging Spectrograph
(IRIS).} {A single Gaussian function is applied to each CO line
profile to extract the line shift, while the moment analysis method
is used for the \mgii~k line. Then the sunspot oscillation can be
found in the time--distance image of Doppler velocities, and the
quasi-periodicity at the sunspot umbra are determined from the
wavelet power spectrum. Finally, the cross-correlation method is
used to analyze the phase relation between different atmospheric
levels.} {At the sunspot umbra, a periodicity of roughly 5~min is
detected at the Doppler velocity range of the CO~7-6~R67 line that
formed in the photosphere, while a periodicity of around 3~min is
discovered at the Doppler velocities of CO~3-2~R14 and \mgii~k lines
that formed in the upper photosphere or the temperature minimum
region and the chromosphere. A time delay of about 2 min is measured
between the strong CO~3-2~R14 line and the \mgii~k line.} {Based on
the spectroscopic observations from the CYRA and IRIS, the 3 min
sunspot oscillation can be spatially resolved in the Doppler shifts.
It may come from the upper photosphere or the temperature minimum
region and then propagate to the chromosphere, which might be
regarded as a propagating slow magnetoacoustic wave.}


\maketitle

\section{Introduction}
Sunspots are  striking features on the solar disk that often exhibit
dark and cool characteristics. They were observed by the naked eye
as early as 2000 years ago \citep[see][]{Wittmann87,Yau88}, and then
were measured by telescope in a white light image
\citep[e.g.,][]{Wolf61,Maunder04}. A typical sunspot is composed of
an umbra characterized by a very dark core and a penumbra that
exhibits a less dark halo;  the sunspot umbra is often separated
into two or more small pieces by one or several bright light bridges
\citep{Sobotka94,Lagg14,Toriumi15,Feng20a}. It is well accepted that
the sunspot is the concentration region of a strong magnetic field
\citep{Cram81,Solanki03,Borrero11}, so convection is strongly
inhibited and further blocks the heat from the solar interior to the
surface, resulting in cool temperatures at the sunspot of
$\sim$4000~K \citep{Rimmele97,Schussler06,Khomenko12,Zhang17}.

Sunspot oscillations can be observed at various layers in solar
atmosphere, and are often interpreted in terms of
magnetohydrodynamic waves
\citep[e.g.,][]{Bogdan00,Chae15,Khomenko15,Jess16,Zhugzhda18,Chae19}.
The dominant period of sunspot oscillations in the low solar
atmosphere (photosphere) is about five minutes
\citep{Beckers72,Lites88,Wang20}, which is believed to be related to
the five-minute p-mode wave
\citep{Thomas85,Bogdan00,Solanki03,Yuan15}. Instead, sunspot
oscillations in the middle solar atmosphere (chromosphere and
transition regions) often have a typical period of around
three~minutes, which can also be observed at photospheric sunspots
\citep{Solanki96,Bogdan00,Yang17}. They are thought to be resonant
modes of sunspot oscillations, with cavities that may be located at
sunspot umbrae in the solar layers of sub-photospheres
\citep{Scheuer81,Thomas84,Thomas85} or chromospheres
\citep{Uexkuell83,Gurman87,Khomenko15}. The typical three-minute
oscillation above the sunspot can be simultaneously detected from
the photosphere through the chromosphere and transition region to
the corona, supporting the interpretation of propagating waves at
sunspots \citep{Dem02,Oshea02,Brynildsen04,Su16,Chae19}, for
instance trans-sunspot waves \citep{Tziotziou06} or slow
magnetoacoustic waves
\citep{Bloomfield07,Krishna15,Wang18,Cho19,Cho20}.

Sunspot oscillations are easily observed at   Doppler velocities. At
photospheric layers the velocity fluctuations at both umbra and
penumbra are similar to their surrounding photosphere
\citep{Khomenko15}. However, the oscillation amplitudes at sunspot
umbra and penumbra are significantly reduced when compared to the
surroundings, i.e., no more than 1 km~s$^{-1}$
\citep{Howard68,Soltau76,Lites88,Chae17,Cho19}. The Doppler shift
oscillations of sunspots are much more apparent and easier to be
detected in chromosphere, such as the spectral lines of \caii~H \&
K, \hei~10830~{\AA}, H$\alpha$ off-band, \mgii~h \& k, and \cii
\citep{Lites86,Solanki96,Bogdan00,Khomenko15}. Moreover, the
oscillation amplitudes at chromospheric umbra are much larger, which
can be about 10~km~s$^{-1}$ or even much larger
\citep{Centeno08,Tian14}. It is still very difficult to observe the
intensity fluctuations at photospheric sunspots since their
amplitudes are very small \citep{Beckers72,Bellot00}. Although some
authors have detected the very weak intensity fluctuations in
G-band, TiO, or visible continuous images
\citep{Nagashima07,Yuan14,Su16},  in the  chromosphere the intensity
oscillations of sunspots are called ``umbral flashes'', which are
often accompanied by the up and down motions with periods of about
2$-$3~min \citep{Wittmann69,Phillis75,Rouppe03,Feng14}.

Part of the fundamental vibration-rotation transition lines of
carbon monoxide (CO) are in the solar infrared spectrum near
46655~{\AA} \citep{Ayres89,Goorvitch94}. The CO molecular lines
contain a wealth of cool plasmas at low solar atmospheres where the
temperature could be as cool as $\sim$3700~K
\citep{Solanki94,Ayres02,Uitenbroek00}. Therefore, they are valuable
tools to investigate the dynamic behavior of the cold heart of the
solar atmosphere, i.e., the temperature minimum region between the
upper photosphere and the lower chromosphere on the Sun
\citep{Ayres06,Penn14}. For example, based on the spatially resolved
solar infrared CO spectrum, a primary 3 min period is found in the
line-center intensity or depth, while the dominant 5 min oscillation
is clearly seen at the Doppler velocity. Both the normal and inverse
Evershed flows at sunspot penumbra are observed
\citep{Uitenbroek94}. According to the study with the McMath/Pierce
solar telescope on Kitt Peak at National Solar Observatory, the
sunspot oscillations at umbra are well separated by double periods
of $\sim$3~min and $\sim$5~min in CO molecular lines
\citep{Solanki96}. Utilizing the same facility, oscillations across
the whole Sun are detected from the line-core intensity and Doppler
velocity in the molecular lines of CO, and these oscillations are
thought to be solar p-modes \citep{Penn11}. On the other hand, the
typical 5 min oscillation in the photospheric layer is also found in
CO molecular lines, and the peak-to-peak amplitudes of
brightness-temperature fluctuations are roughly 225$-$300~K, while
the peak-to-peak amplitudes at Doppler velocities are
$\sim$1.1~km~s$^{-1}$ \citep{Noyes72,Ayres90}.

In this paper, we investigated the sunspot oscillation using the
solar infrared spectrum and near-ultraviolet (NUV) line, i.e., the
CO molecular lines (3-2 R14 and 7-6 R67) and \mgii~k line. We
focused on the Doppler shift oscillations at the umbra. Our data are
compiled from  ground- and space-based observations, such as the
Cryogenic Infrared Spectrograph \citep[CYRA,][]{Cao10,Cao12} at Big
Bear Solar Observatory (BBSO), and the Interface Region Imaging
Spectrograph \citep[IRIS,][]{Dep14}.

\section{Observations}

\begin{table*}
\caption{Details of observational instruments presented in this paper.}
\centering \setlength{\tabcolsep}{15pt}
\begin{tabular}{c c c c c c c}
 \hline\hline
Instruments &  Channels        &  Cadence     &  Spectral dispersion             &     Pixel size       \\
\hline
            &  CO 3-2 R14      & $\sim$15~s   &  $\sim$34.85~m{\AA}~pixel$^{-1}$ &  $\sim$0.16\arcsec   \\
CYRA        &  CO 7-6 R67      & $\sim$15~s   &  $\sim$34.85~m{\AA}~pixel$^{-1}$ &  $\sim$0.16\arcsec   \\
\hline
            & \mgii~k          & $\sim$19~s   &  $\sim$25.46~m{\AA}~pixel$^{-1}$ &  $\sim$0.166\arcsec   \\
IRIS        &  SJI~2796~{\AA}  & $\sim$37~s   &            -                     &  $\sim$0.166\arcsec   \\
            &  SJI~2832~{\AA}  & $\sim$223~s  &            -                     &  $\sim$0.166\arcsec   \\
 \hline\hline
\end{tabular}
\label{tab1}
\end{table*}

\begin{figure}
\centering
\includegraphics[width=\linewidth,clip=]{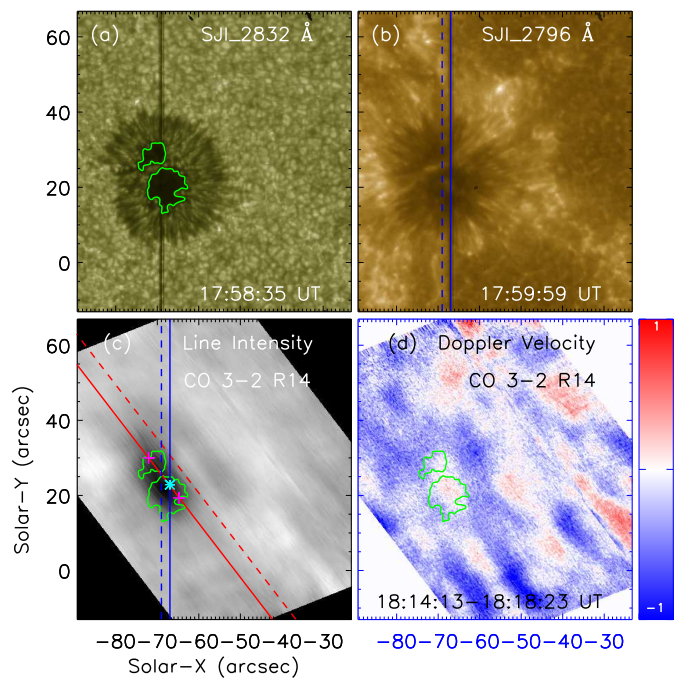}
\caption{Near-UV snapshots and CO large scan images on 2017
September 15 measured by the IRIS/SJI and GST/CYRA. The green
contours represent the umbra--penumbra boundary. Two blue lines
outline IRIS slits, and double red lines mark the first and last
(13th) slits of CYRA. The two magenta plus signs ($+$)  in panel~(c)
indicate the umbra--penumbra boundary positions at the first CYRA
slit. The cyan asterisk ($\ast$) in panel~(c) indicates a crossover
point between the slits of IRIS and CYRA at the sunspot umbra.
\label{snap}}
\end{figure}

On 2017 September 15, a sunspot near the solar disk center at the
active region of NOAA 12680 (N08E01) was measured by the ground- and
space-based telescopes listed in Table~\ref{tab1}.
Figure~\ref{snap}~(a)$-$(b) shows the NUV images with a field of
view (FOV) of about 66\arcsec$\times$80\arcsec\ in Slit-Jaw Imager
(SJI)~2832~{\AA} and 2796~{\AA} on board IRIS. SJI~2796~{\AA} image
contains radiation primarily from the \mgii~k line, which is formed
in the chromosphere with a formation temperature of $\sim$10$^4$~K.
SJI~2832~{\AA} emits the radiation dominated by \mgii wing, where
the formation temperature is roughly (5$-$8)$\times$10$^3$~K
\citep{Dep14}. A typical sunspot near solar disk center can be seen
in panel~(a), which is composed of two umbrae and one penumbra;  the
umbrae are separated into two pieces by a light bridge, as outlined
by the green contours. In this observation, IRIS scans the sunspot
in a ``two-step raster'' mode from about 16:40:15~UT to 18:01:59~UT.
The step size is $\sim$2\arcsec, and the time cadence between two
nearby scans is roughly 19~s. The double slits of IRIS are along the
solar north--south direction and go through the sunspot,
particularly scanning the umbrae, as indicated by the two blue
vertical lines in panel~(b).

The sunspot was also measured by the CYRA, which is installed on the
1.6~m Goode Solar Telescope (GST) at BBSO
\citep{Cao10,Cao12,Yang20}. Several CO molecular bands spread
throughout the CYRA spectrum, and the fundamental rotation-vibration
transition lines near 46655~{\AA} are unique because they are not
obscured by the Earth's atmosphere. Therefore, they are well-known
diagnostics of the lower atmosphere of the Sun
\citep[e.g.,][]{Solanki94,Cao10,Penn14}. During our observations
GST/CYRA measured the sunspot between $\sim$16:50:14~UT and
$\sim$17:15:14~UT, and it performed small quick 13-step raster scans
with a step size of $\sim$0.4\arcsec, while the time cadence between
two nearby raster scans is about 15~s. To co-align with other
instruments, we also performed a large 400-step raster scan between
around 18:14:13$-$18:18:23~UT with a step size of $\sim$0.2\arcsec.
Figure~\ref{snap}~(c)$-$(d) present the large scan images with the
same FOV of around 66\arcsec$\times$80\arcsec\ in the line intensity
and at the Doppler velocity of the CO~3-2~R14 line, respectively.
Two oblique red lines indicate the first (solid) and 13th (dashed)
slits of CYRA small raster scan, which are along a
$\sim$29$^{\circ}$ angle to the solar north--south direction. Two
magenta crosses  indicate the crossover points between the first
CYRA slit and the umbra--penumbra boundary. It can be seen that the
slits of CYRA and IRIS overlap in one position
($x~\approx$~-67\arcsec, $y~\approx$~23\arcsec) at the sunspot
umbra, as shown by the cyan asterisk.

\begin{figure}
\centering
\includegraphics[width=\linewidth,clip=]{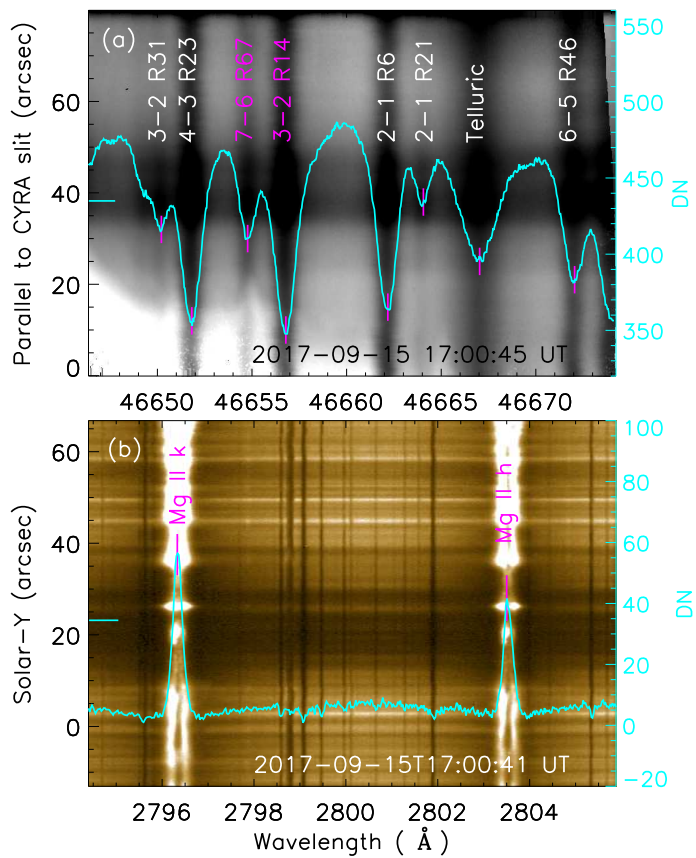}
\caption{Solar spectra in infrared (a) and NUV (b) wavebands at
around $\sim$17:00~UT measured by CYRA and IRIS, respectively. The
overplotted curves are the line spectra marked by a cyan line on
the left side of each image. The main lines are labeled and
indicated by magenta vertical ticks. \label{spect}}
\end{figure}

Figure~\ref{spect}~(a) shows the solar infrared CO spectrum at
$\sim$17:00:45~UT observed by GST/CYRA in its first slit, as
indicated by the red solid line in Figure~\ref{snap}. It has been
pre-processed (see details in Appendix~\ref{instrument}). We note
that the bright patch in the lower left corner of
Figure~\ref{spect}~(a) are the hot pixels from the detector, which
cannot be effectively corrected by the interpolation due to their
patchy distribution. So we avoid using these regions in the
following analysis. The overplotted curve in panel~(a) is the CO
line profile at the slit position of $\sim$38\arcsec, as indicated
by a short cyan line on the left side. A number of absorption lines
can be seen from the solar infrared spectrum, and eight obvious
lines are identified and labeled with short magenta ticks, including
seven CO molecular lines and a telluric line. In this paper, two CO
molecular lines are used to study the sunspot oscillation: a strong
line of CO 3-2~R14 and a weak line of CO~7-6~R67. Their formation
height is around 500~km above the $\tau_{500}$~=~1 (z~=~0), ranging
from the photosphere through temperature minimum region to the low
chromosphere. The weak line (CO 7-6 R67) is probing the lower solar
layers, such as the photosphere. While the strong line (CO 3-2 R14)
could vary from roughly 150 to 580~km above $\tau_{500}$~=~1, which
probes the upper photosphere, the temperature minimum region, as
well as the lower chromosphere \citep[see
also][]{Uitenbroek00,Ayres06}.

Figure~\ref{spect}~(b) presents the IRIS spectrum in NUV wavebands
at around 17:00:41~UT in the second slit, as indicated by the blue
solid line in Figure~\ref{snap}. They have been pre-processed with
the standard IRIS routines in SSW package \citep{Dep14}. The
overplotted curves are the line profiles at the solar position
nearby $y\approx$~23\arcsec, as marked by a short cyan line on the
left side of each panel. The double resonance lines of \mgii~k \& h
are identified in panel~(b), and they are mostly formed in the
chromosphere with a formation temperature of $\sim$10$^4$~K.

\section{Data analysis and results}
\subsection{Spectroscopic diagnostics with two CO lines}
GST/CYRA observed the sunspot in a quickly raster mode with 13
steps, and the scanned FOV was about 4.8\arcsec$\times$80\arcsec,
while the exposure time was $\sim$40~ms. The 13 slits of CYRA went
through the sunspot umbra and penumbra in sequence, as shown in
Figure~\ref{snap}~(c). Figure~\ref{spect} suggests that each CO
molecular line near its center is clearly a  Gaussian profile. Thus,
we applied a single Gaussian fit (see Appendix~\ref{fit-co}) to each
line profile of CO molecule, i.e., CO 3-2 R14 and CO 7-6 R67
\citep[see][]{Uitenbroek94,Uitenbroek00}. Then their Doppler
velocities and line intensities could be measured.

\begin{figure}
\centering
\includegraphics[width=\linewidth,clip=]{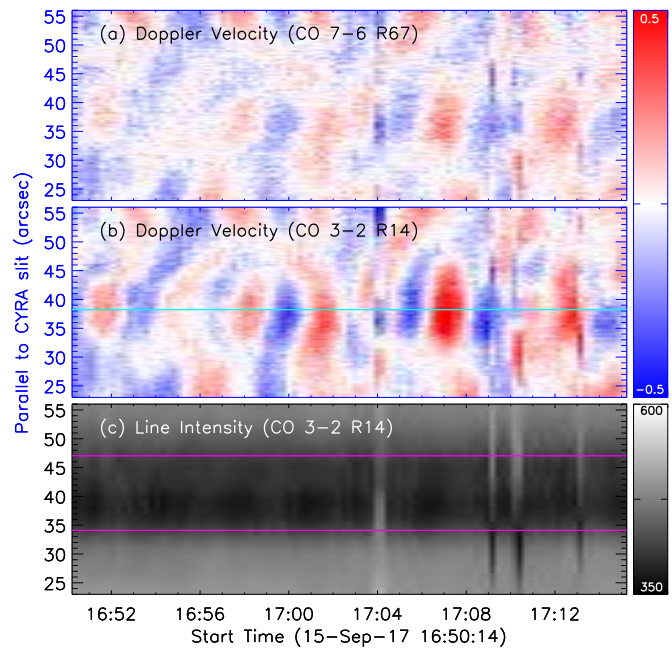}
\caption{Time--distance images along the first CYRA slit (red solid
line in Figure~\ref{snap}) of Doppler velocity and intensity in
CO~3-2~R14 and CO~7-6~R67 lines. The $y$-axis is parallel to the
slits of CYRA. A horizontal cyan line in panel~(b) outline the
umbral position to perform the wavelet analysis in Figure~\ref{wav1}
and \ref{wav2}. Two magenta lines outline the umbra--penumbra
boundary. \label{covy}}
\end{figure}

Figure~\ref{covy} presents the time--distance images derived from
the first slit of CYRA. That is, the $y$-axis is parallel to the
slit direction between about 23\arcsec and 56\arcsec. We note that
the short time disturbances near $\sim$17:04~UT and $\sim$17:10~UT
correspond to the time when the image stabilization system does not
work. It can be seen from panels~(a) and (b) that the Doppler
velocities of CO molecular lines are small, roughly
$\pm$0.5~km~s$^{-1}$. They all exhibit a pronounced signature of
oscillations that change from redshifts to blueshifts, in particular
at the umbra, i.e., between 35\arcsec\ and 40\arcsec. The
oscillation periods can be determined from the numbers of red
patterns. Then a long period of roughly 5~min can be found in the
CO~7-6~R67 line, while a short period of around 3~min can be
discovered in the CO~3-2 R14 line at the sunspot umbra. We  also
find that the Doppler shift oscillations are similar at the
penumbra, i.e., at the slit position near 30\arcsec and 50\arcsec.
However, there is not any apparent oscillation in the intensity
image (c) of CO~3-2~R14 line at the sunspot. Two magenta lines mark
the umbra--penumbra boundaries, which are indicated by the two plus
signs ($+$) in Figure~\ref{snap}~(c). Here we do not show the line
intensity of CO~7-6~R67 line since it does not exhibit the
pronounced oscillation feature.

\begin{figure}
\centering
\includegraphics[width=\linewidth,clip=]{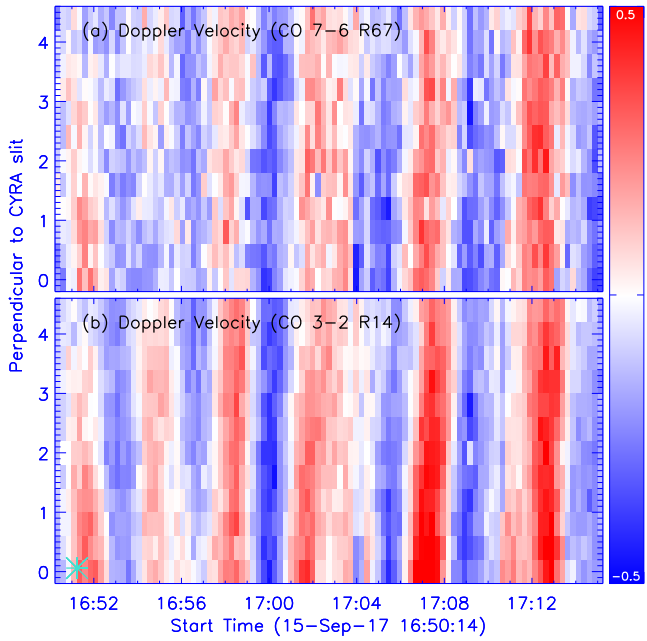}
\caption{Time--distance images of Doppler velocity in CO~3-2~R14 and
CO~7-6~R67 lines. The $y$-axis is perpendicular to the slits of
CYRA. The cyan asterisks ($\ast$) symbol indicates the start point
on the $y$-axis, which is same as in Figure~\ref{snap}~(c).
\label{covx}}
\end{figure}

To spatially resolved  the sunspot oscillation, we then plot the
time--distance images along the CYRA scanned direction, as shown in
Figure~\ref{covx}. In other words, the $y$-axis is perpendicular to
the slits of CYRA with a length of around 4.8\arcsec, and the first
slit corresponds to bottom position, while the 13th slit is in the
upper location. Their Doppler velocities are both characterized by a
number of vertical slashes that change from redshift to blueshift,
including the strong line of CO~3-2~R14 and the weak CO~7-6~R67
line. These repeating dynamical behaviors can be considered  sunspot
oscillations. It seems that the CO~3-2~R14 and CO~7-6~R67 lines both
show the same period of 5 min at most regions. In the bottom region
($\ast$) of panel~(b), the there seem to be more red patterns than
in the other regions, suggesting a short period at the umbra. All
these observational results agree with the previous findings (see
Figure~\ref{covy}).

\subsection{Spectroscopic observations in the \mgii~k line}
IRIS measured the sunspot in two-step raster mode between
16:40:15~UT and 18:01:59~UT. The spectrograph aboard IRIS covered a
small FOV of around 2\arcsec$\times$119\arcsec, with an exposure
time of $\sim$8~s. Luckily, the slits of IRIS nearly crossed  the
sunspot center, and the second slit (blue solid line in
Figure~\ref{snap}) was crossed by the first slit of CYRA (red solid
line in Figure~\ref{snap}) at the sunspot umbra, (cyan asterisk in
Figure~\ref{snap}~(g)). So, the IRIS spectra from its second slit
were used to investigate the sunspot oscillation. The double
resonance lines of \mgii~k~$\&$~h are usually optically thick in
solar spectra, and their line profiles often exhibit central
reversals at line cores \citep{Leenaarts13,Cheng15}. However, the
\mgii~k \& h lines at sunspots do not appear to be the prominent
central reversals of line cores \citep[see][]{Tian14,Zhang17}.
Figure~\ref{spect}~(b) shows the line profiles of \mgii~k \& h lines
at sunspot umbra, and their line cores are not central reversals,
but they are also non-Gaussian profiles. Therefore, the moment
analysis method but not the single Gaussian fit is applied to
estimate their Doppler velocities, line widths, and intensities
\citep[see detail in][]{Li17}.

\begin{figure}
\centering
\includegraphics[width=\linewidth,clip=]{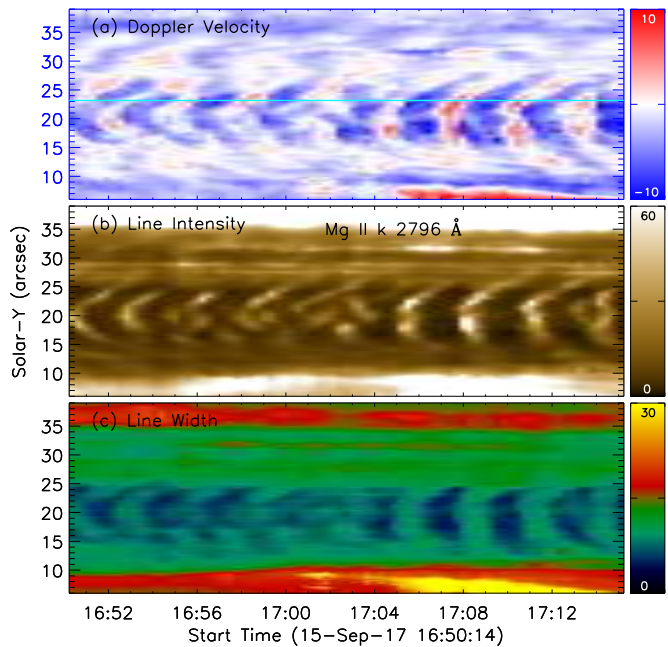}
\caption{Time--distance images along the second slit of IRIS (blue
solid line in Figure~\ref{snap}) of Doppler velocity, line width, and
intensity in \mgii~k. The horizontal cyan line in panel~(a) gives
the umbral position, used to perform the wavelet analysis in
Figure~\ref{wav3}. \label{mgvy}}
\end{figure}

Figure~\ref{mgvy} presents the time--distance images of Doppler
velocity (a), line intensity (b), and line width (c) in the \mgii~k
line. Here, the $y$-axis is parallel to the slits of IRIS (blue
lines) with a length of $\sim$33\arcsec\ in the approximate range of
6\arcsec$-$39\arcsec\ along the solar-Y direction. The Doppler
velocities of \mgii~k line at the umbra near $y\sim$~23\arcsec\
exhibit pronounced oscillations with a quasi-period of around
3~min. They are characterized by a group of repeating oblique
streaks, which  changing from redshift to blueshift. The
oblique streaks suggest that they are propagating and eventually
become invisible at the  umbra--penumbra  boundary. The line
intensity and width show the same oscillation behaviors. For
example, the 3 min oscillation is identified as the repeating
oblique slashes at the sunspot umbra, which also exhibit a
propagation movement and gradually disappear when they reach the
outer umbral boundary. Finally, we do not find any clear signature
of oscillations at the penumbra, such as at the positions of around
$y\sim$~10\arcsec\ and $y\sim$~35\arcsec.

\subsection{Wavelet analysis}
To look closely at the periods of the sunspot oscillation at the
umbra, we then performed wavelet analysis
\citep[e.g.,][]{Torrence98,Chae17,Cho19} on the detrended time
series of Doppler velocities at the sunspot umbra, as indicated by
the cyan line in Figures~\ref{covy} and \ref{mgvy}. It should be
noted that it is the same position at the sunspot umbra, which is
indicated with the cyan asterisk in Figure~\ref{snap}~(c). The
detrended time series are used here because we thereby enhance the
periods, for example 3 or 5  min. The discussion and application of
this method can be found in previous works
\citep{Gruber11,Kupriyanova13,Li20}.

\begin{figure}
\centering
\includegraphics[width=\linewidth,clip=]{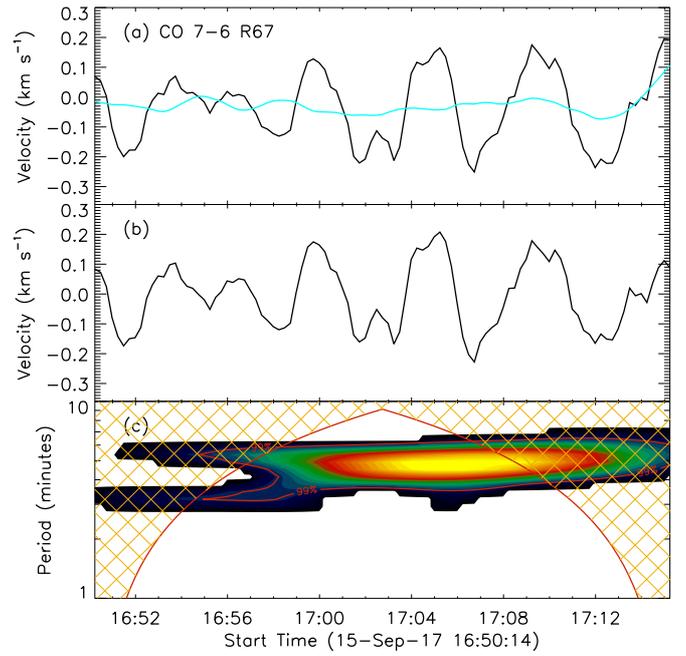}
\caption{Wavelet analysis result in the CO~7-6~R67 line at the sunspot
umbra. Panel~(a): Doppler velocity (black) and its trend (cyan).
Panel~(b): Detrended velocity. Panel~(c): Wavelet power spectrum.
The red line indicates a significance level of 99\%. \label{wav1}}
\end{figure}

\begin{figure}
\centering
\includegraphics[width=\linewidth,clip=0]{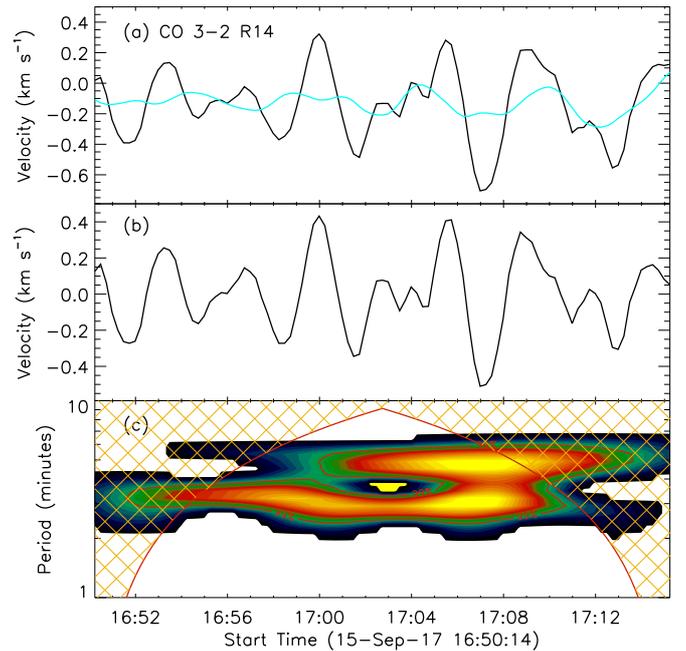}
\caption{Wavelet analysis result in the CO~3-2~R14 line at the same
umbral position. Panel~(a): Doppler velocity (black) and its trend
(cyan). Panel~(b): Detrended velocity. Panel~(c): Wavelet power
spectrum. The red line indicates a significance level of 99\%.
\label{wav2}}
\end{figure}

Figure~\ref{wav1} presents the wavelet analysis result of the
Doppler velocity in CO~7-6~R67 line. Panel~(a) shows the time series
between $\sim$16:50:14~UT and 17:15:14~UT of the Doppler velocity at
the sunspot umbra, as indicated with the cyan asterisk  in
Figure~\ref{snap}~(c). The Doppler velocity (black) in the
CO~7-6~R67 line is very low, with a peak value of roughly
0.2~km~s$^{-1}$. Then the trend Doppler velocity is overplotted with
cyan line, which is a four-minute running average. It can be seen
that the trend velocity is close to zero. Panel~(b) displays the
detrended velocity after subtracting the four-minute running average
\citep[e.g.,][]{Wang09,Li18}, which exhibits five pronounced peaks
within a duration of 25~minutes, suggesting a 5 min periodicity. The
5 min periodicity is confirmed by the wavelet power spectrum shown
in panel~(c). It clearly shows a period of 5~min above the
confidence level of 99\% (red contour).

Figure~\ref{wav2} shows the wavelet analysis result of the Doppler
velocity in CO~3-2~R14 line during the same time interval, i.e.,
from $\sim$16:50:14~UT to $\sim$17:15:14~UT. The Doppler velocity
here is a little higher than that of CO~7-6~R67 line, i.e., the peak
value can be reach to $\sim$0.5~km~s$^{-1}$, as shown by the black
curve in panel~(a). The detrended velocity also appears as several
pronounced peaks, but there are also some small peaks, as seen in
panel~(b). Thus, it is hard to determine a single period through the
peaks. The period can be identified in the wavelet power spectrum,
as given in panel~(c). It exhibits two prominent periods, one is a 3
min periodicity and can be detected almost at the observed time, the
other is 5 min periodicity that only appears after $\sim$17:00~UT.

\begin{figure}
\centering
\includegraphics[width=\linewidth,clip=0]{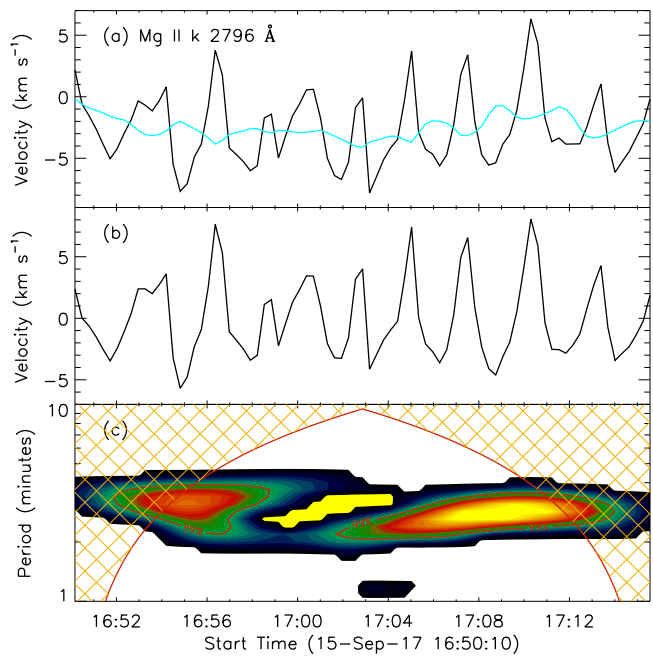}
\caption{Wavelet analysis result in the \mgii~k line at the same umbral
position. Panel~(a): Doppler velocity (black) and its trend
(cyan). Panel~(b): Detrended velocity. Panel~(c): Wavelet power
spectrum. The red line indicates a significance level of 99\%.
\label{wav3}}
\end{figure}

The similar wavelet analysis result of the Doppler velocity in
\mgii~k line is shown in Figure~\ref{wav3}. At the same umbra
position, The Doppler velocities are much higher that those in the
CO molecular lines, which can be as high as $\sim$8~km~s$^{-1}$, as
shown in panel~(a). The wavelet power spectrum exhibits a dominant
period of around 3 min, which agrees with the 3 min periodicity
derived from the CO~3-2~R14 line.

\subsection{Cross-correlation analysis}
\begin{figure}
\centering
\includegraphics[width=\linewidth,clip=0]{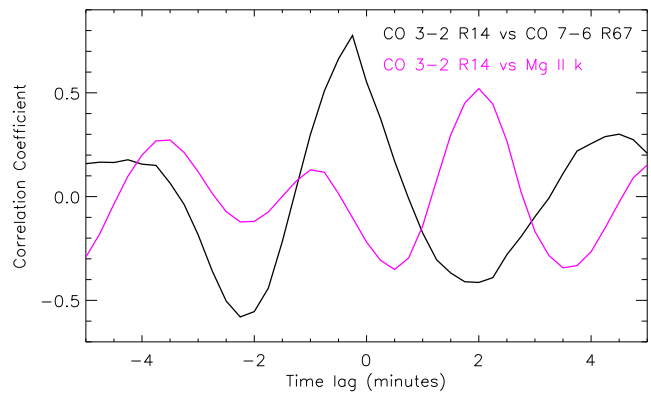}
\caption{Correlation coefficients between two parameters as a
function of the time lag: the Doppler velocities between two CO
molecular lines (black), and the strong CO~3-2~R14 and \mgii~k lines
(magenta). \label{tlag}}
\end{figure}

To analyze the phase relation between different atmospheric levels,
for example two CO molecular lines and \mgii~k line, a
cross-correlation analysis \citep[e.g.,][]{Tian14,Krishna15,Su16} is
applied for the detrended time series to investigate their time
delays, as shown in Figure~\ref{tlag}. A maximum correlation
coefficient of $\sim$0.77 between the strong CO~3-2~R14 line and the
weak CO~7-6~R67 line is found at the time lag of around -0.3~minute,
implying a short time delay between them; in other words, the
formation height of the strong CO~3-2~R14 line is a little higher
than that of the weak CO~7-6~R67 line, which is similar to previous
observations \citep{Uitenbroek00,Ayres02}. On the other hand, a
maximum correlation coefficient of $\sim$0.52 between CO~3-2~R14 and
\mgii~k lines is discovered at the time lag of about 2~min,
suggesting a long time delay between them. It should be noted  that
the time delay between CO~3-2~R14 and \mgii~k lines here suggests
that the 3 min oscillation at sunspot umbra is a propagating wave,
which agrees with previous observations about the propagating slow
wave in sunspots \citep[see][]{Khomenko15,Krishna15,Wang18}.

\section{Conclusion and discussion}
Using the spectroscopic observations in solar infrared and NUV bands
measured by the GST/CYRA and the IRIS, we investigated the sunspot
oscillation in Doppler velocities of two CO molecular lines and the
\mgii~k line. The weak line of CO 7-6 R67 exhibits the 5 min
oscillation at the sunspot umbra (Figure~\ref{wav1}). However, the
strong CO 3-2 R14 line shows double periods at the sunspot umbra
(Figure~\ref{wav2}) of nearly 5~min and roughly 3~min. The double
oscillation periods could be due to the fact that the strong CO 3-2
R14 line contains two-layer radiation on the Sun, i.e., the upper
photosphere and temperature minimum region
\citep{Uitenbroek00,Ayres06}. The line profiles of \mgii~k lines
also show the typical 3 min oscillation at the sunspot umbra, as
shown in Figure~\ref{wav3}. On the other hand, an oscillation period
of around 5~min can be found in the two CO molecular lines at the
sunspot penumbra, but it is not detected in the \mgii~k line.
Therefore, only the sunspot oscillation at the umbra is studied in
detail.

An interesting aspect is that the sunspot oscillation can be
spatially resolved in the Doppler shifts, i.e., they can be
discovered in three directions above sunspot at the solar
atmosphere, for instance the directions that are perpendicular and
parallel to the CYRA slit, and also parallel to the IRIS slit. This
is different from previous spatial distributions of sunspot
oscillations observed in intensity images
\citep[e.g.,][]{Yuan14,Wang20,Yurchyshyn20}. Figures~\ref{covy} and
\ref{covx} demonstrate that the sunspot oscillations at the umbra
and penumbra can appear in the directions perpendicular and parallel
to the CYRA slits, including the long period of nearly 5~min and the
short period of around 3~min. Meanwhile, we demonstrated that the
CYRA slits show a roll angle of $\sim$29$^{\circ}$ with the slits of
IRIS, as can be seen in Figure~\ref{snap}~(c). The 3 min oscillation
at the umbra (around $y\sim$~20\arcsec) can also be found parallel
to the slits of IRIS, as can be seen in Figures~\ref{mgvy}.
Therefore, our observational results suggest that the sunspot
oscillation (particularly the umbral oscillation) can be found in an
arbitrary direction, which is similar to the running umbral waves
\citep{Alissandrakis98,Kobanov04}. All these observational results
imply that they can be considered trans-sunspot waves
\citep{Tziotziou06,Chae17}. Finally, the sunspot oscillation could
be detected in two perpendicular directions, which is benefited from
the fast scan of the CYRA with 13 steps.

We wanted to discuss the observed periods at sunspot. A period of
nearly five~minutes can be detected in Doppler velocities of CO
3-2~R14 and 7-6~R67 lines, which is consistent with previous
findings in white light images or continuum spectrum
\citep{Beckers72,Lites88,Nagashima07,Yuan14,Su16}, and might be
considered the solar p-mode waves in the photosphere
\citep{Thomas85,Bogdan00,Solanki03}. While a period of roughly
three~minutes is found in the Doppler velocities of the CO 3-2~R14
and \mgii~k lines, which agrees closely with the previous
observational results in UV--infrared lines or images at the sunspot
umbra
\citep[e.g.,][]{Solanki96,Bogdan00,Fludra01,Maltby01,Centeno08,Tian14,Khomenko15,Yang17}.
They are explained as the resonant modes of sunspot oscillations
\citep{Uexkuell83,Thomas84,Gurman87,Khomenko15}. On the other hand,
the CO~3-2~R14 line is believed to provide the information in the
upper photosphere or the temperature minimum region
\citep{Uitenbroek00,Ayres02,Ayres06}. Moreover, a time delay of
about two minutes is measured between the CO~3-2~R14 and \mgii~k
lines, as shown in Figure~\ref{tlag}. So, the three-minute
oscillation at the sunspot umbra could come from the upper
photosphere or the temperature minimum region and then propagate to
the chromosphere, supporting the interpretation that propagating
waves above the sunspots originate from the lower solar atmosphere
\citep{Dem02,Oshea02,Brynildsen04,Khomenko15,Krishna15}. In other
words, the three-minute oscillation can be regarded as the upwardly
propagating slow magnetoacoustic waves
\citep[e.g.,][]{Su16,Chae19,Cho20,Feng20}. Finally, \cite{Chae17}
found that the three-minute oscillation in the light bridge or
umbral dots of a sunspot originates from the photosphere \cite[see
aslo][]{Cho19}, which is consistent with our results and further
suggests that the CYRA data is reliable.

\begin{acknowledgements}
We acknowledge the anonymous referee for his/her valuable comments.
BBSO operation is supported by US NSF AGS 1821294 grant and NJIT.
GST operation is partly supported by the Korea Astronomy and Space
Science Institute, the Seoul National University, and the Key
Laboratory of Solar Activities of Chinese Academy of Sciences (CAS)
and the Operation, Maintenance and Upgrading Fund of CAS for
Astronomical Telescopes and Facility Instruments. We also thank the
teams of IRIS for their open data use policy. This study is
supported by NSFC under grants 11973092, 11427901, 11773038,
11873062, 11790300, 11790302, 11729301, 11873095, the Youth Fund of
Jiangsu No. BK20171108, as well as National Natural Science
Foundation of China (U1731241), the Strategic Priority Research
Program on Space Science, CAS, Grant No. XDA15052200 and
XDA15320301. D. Li is supported by CAS Key Laboratory of Solar
Activity (KLSA202003) and the Specialized Research Fund for State
Key Laboratories. The Laboratory No. 2010DP173032.
\end{acknowledgements}

\clearpage
\begin{appendix}
\section{CYRA instrument and data calibration}
\label{instrument} CYRA is the first fully cold cryogenic solar
spectrograph, and the detector currently used by CYRA is a
commercial detector rather than a scientific one \citep{Cao12}. All
CYRA components (e.g., slit, grating, collimator/imager, order sorting
filter, and detector) are placed in a dual-layer cryostat
working at very low temperature, i.e., $\sim$77~K for the outer case
and $\sim$30~K for the inner case. Thus, the thermal background
emission is tremendously minimized \citep{Cao10,Yang20}. It works at
the wavelength region in the range 10000$-$50000~{\AA,} and the maximum
frame rate of the 2~k$\times$2~k detector is 76~Hz.

The raw images observed by the CYRA have been pre-processed in the
following way:
\begin{enumerate}
\item The dark images obtained by exposing with the main cover of
GST closed are subtracted.

\item The dark subtracted images are corrected for flat fields, which
are performed by averaging hundreds of frames taken by randomly
moving the telescope near the solar disk center. The spatially averaged
spectrum along the slit is also removed during the calculation of
flat fields.

\item The values of bad pixels from the dark-subtracted and
flat-fielding images are replaced with a linear interpolation from
neighboring pixels.

\item The spectrum is corrected for the slant and distortion in both
slit and dispersion directions. Then the wavelength jitter or drift
are done according to a telluric line nearby the CO lines. After the
above-mentioned calibration, there are still some residual distinct
stripe patterns, and the derived physical parameters vary along the
slit, which are further obtained and corrected by averaging the
scanned quiet regions in the 400-step raster scanned image and the
time sequences of the derived physical parameters in the 13-step
raster scans.

\item The Doppler velocity calibration is done based on the
phenomenon that the average value of Doppler velocities in the
sunspot umbra is almost zero \citep[e.g.,][]{Lohner18}. The velocity
calibration for the data from the 13-step raster is done with the
assumption that the averaged value in the time sequences is near
zero.
\end{enumerate}

\section{Gaussian fit to the CO line}
\label{fit-co} 
\begin{figure}
\centering
\includegraphics[width=\linewidth]{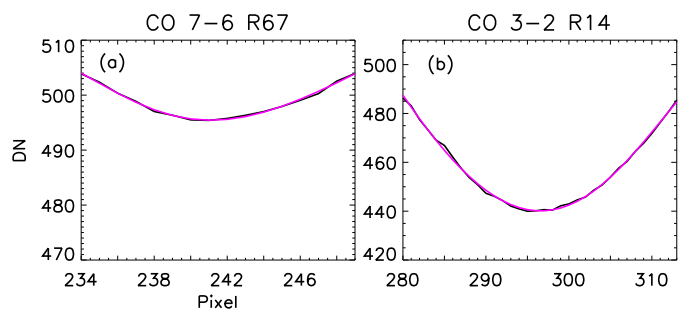}
\caption{Example of the CO weak (a) and strong (b) line fitting
results (magenta) with the single Gaussian fit method. The black
line represents the corresponding observational profile measured by
CYRA. \label{co_fit}}
\end{figure}

In this study we presented the single Gaussian
fitting result and its original observational line profile near the
line center of both the CO 7-6 R67 and the CO 3-2 R14 lines, as
shown in Figure~\ref{co_fit}. It can be found that the single
Gaussian fit agrees well with its corresponding original profile.
Moreover, \cite{Uitenbroek00} also used the single Gaussian fit to
derive these parameters for CO lines. \cite{Penn11} compared the
Gaussian fitting result with Voigt fitting method and no systematic
differences were seen in the derived line center wavelength. On the
basis of the above considerations, we finally used the single
Gaussian fits to the CO lines profiles.

\end{appendix}

\end{document}